\documentstyle[12pt]{article}
\begin{document}
\thispagestyle{empty}
\begin{center}
\LARGE \tt \bf{Domain walls and torsion potentials.}
\end{center}
\vspace{2.5cm}
\begin{center}{\large L.C. Garcia de Andrade\footnote{Departamento de F\'{\i}sica Teorica - UERJ.
Rua S\~{a}o Fco. Xavier 524, Rio de Janeiro, RJ
Maracan\~{a}, CEP:20550-003 , Brasil.
{E-Mail.: GARCIA@SYMBCOMP.UERJ.BR}}}
\end{center}
\vspace{2.0cm}
\begin{abstract}
A Lagrangian for flat domain walls in spaces with Cartan torsion and electromagnetic 
fields is proposed.The Lagrangian is very similar to a recently proposed Lagrangian for 
domain walls in a Chern-Simons electrodynamics in 2+1 dimensions.We show that in the 
first approximation of the torsion scalar potential the field equations are reduced to 
a Klein-Gordon type field equation for the torsion potential and the electromagnetic 
wave equation.A planar symmetric solution representing a parallel plates electric 
capacitor interacting with the electric field is given.The photon mass is proportional to the torsion potential and in the time dependent case the angular momentum is computed and is shown to be connected with torsion in analogy with the spin-torsion relation which appears in Einstein-Cartan gravity.When the curvature Ricci scalar is introduced we are able to show that the torsion potential can be associated with Higgs massive vectorial bosons.
\end{abstract}
\vspace{2.0cm}
\newpage
Recently vortices and domain walls in a Chern-Simons (CS) theory 
with magnetic moment interactions were investigated by Antillon,
Escalona and Torres \cite{1}.In their paper they work in 2+1 
dimensional spacetime.Here following recently investigation on 
non-Riemannian domain walls in space-times with torsion \cite{2,3,4} we worked out avery similar Lagrangian like the one 
of reference one with the exception that dual field in the (CS) 
Lagrangian is replaced by torsion and we work in the full 
four-dimensional spacetime.The equations obtained for the torsion 
potential and the electromagnetic field are reduced to non-linear 
Klein-Gordon field equation and to the Proca field equation in the 
case of the electromagnetic vector potetencial.The controversy 
concerning the interaction of the electromagnetic fields and 
torsion \cite{5,6,7} here reachs a slightly different approach 
since the interaction between photons and torsion appears 
intermediated by a spin-0 massive boson given by the torsion 
potential ,besides in the first approximation of the torsion 
potential the Proca equation is reduced to the Maxwell equation 
and the photon is massless and the theory is gauge invariant in 
first approximation.As an application we solve the field equations 
in the first order in the torsion potential for the case of the 
parallel plates Capacitor.This simple example is analogous to the 
case of the Casimir effect \cite{8} with torsion to be investigated 
in a future research.Of course throughout the paper the metric is 
considered to be flat.Let us assume that our Lagrangian is given 
by
\begin{equation}
L=-\frac{1}{4}F^{2}+\frac{1}{2}|D_{\mu}{\phi}|^{2}-V({\phi})
\label{1}
\end{equation}
where the bars denotes the modulus since we considering that the 
torsion potential ${\phi}$ is real but the covariant derivative 
defined by 
\begin{equation}
D_{\mu}={\partial}_{\mu}-ifA_{\mu}-igS_{\mu} 
\label{2}
\end{equation}
is complex.Here $F^{2}=F_{\mu\nu}F^{\mu\nu}$ is the electromagnetic 
invariant and $F_{\mu\nu}={\partial}_{\mu}A_{\nu}-{\partial}_{\nu}A_{\mu}$ 
is the electromagnetic field tensor.The torsion potential 
generates the torsion vector through the relation 
$S_{\mu}={\partial}_{\mu}{\phi}$.Substitution of the definition 
(\ref{2}) into the Lagrangian (\ref{1}) reduces the previous Lagrangian to 
\begin{equation}
L=\frac{1}{2}({\partial}_{\mu}{\phi})^{2}-\frac{1}{4}F^{2}-\frac{1}{2}(f^{2}A^{2}+g^{2}S^{2}){\phi}^{2}-V({\phi})
\label{3}
\end{equation}
or
\begin{equation}
L=\frac{1}{2}(1+f^{2})({\partial}{\phi})^{2}+\frac{1}{2}g^{2}A^{2}{\phi}^{2}-\frac{1}{4}F^{2}-V({\phi})
\label{4}
\end{equation}
Variation of this Lagrangian with respect to torsion potential and 
the electromagnetic potential vector $A_{\mu}$ yields the following 
non-linear field equations
\begin{equation}
{\nabla}^{2}A^{\mu}+\frac{1}{2}g^{2}{\phi}^{2}A^{\mu}=0
\label{5}
\end{equation}
and 
\begin{equation}
{\partial}^{2}{\phi}-\frac{2f^{2}{\phi}}{(1+f^{2}{\phi}^{2})}
({\partial}{\phi})^{2}=\frac{g^{2}A^{2}{\phi}}{1+f^{2}{\phi}^{2}}
\label{6}
\end{equation}
where we have supressed the indices and have already considered the 
domain wall potential $V({\phi})$ as zero since we will not need 
it in our next application.Notice that the equation (\ref{5}) yields
a mass for the photon like $ m_{\gamma}=g{\phi} $.Thus the photon 
mass can be expressed in terms of the torsion potential.This situation is
similar to the photon mass dependence on torsion developed some 
years ago by Sivaram and myself \cite{9}.One should notice that equation 
(\ref{4}) is the Proca equation for the massive photon while the 
second is a non-linear field equation which reduces to the 
Klein-Gordon type field equation in the case that the terms 
quadratic in the torsion potential are dropped out.In this case 
also we note that the Proca equation is reduced to the massless 
wave equation which in the time independent case reads
\begin{equation}
{\nabla}^{2}A^{\mu}=0
\label{7}
\end{equation}
As an example let us solve these equations inthe static case of a 
parallel plate capacitor interacting with the scalar torsion field.
Since the capacitor electrostatic potential is given by ${\phi}^{el}=qd $ where d is the separation between the plates and q is the electric charge.Here we have taken the Parallel plate orthogonal to the z-coordinate axis,the Klein-Gordon-Cartan 
equation is
\begin{equation}
{\phi}"-g^{2}d^{2}q^{2}{\phi}=0
\label{8}
\end{equation}
where the double lines denote second derivative in $z$.To solve this 
Klein-Gordon field equation for torsion potential,let us use the 
following ansatz
\begin{equation}
{\phi}=e^{-{\alpha}z}
\label{9}
\end{equation}
Since we assume that the torsion field vanishes at infinity.
Substitution of (\ref{9}) into (\ref{8}) yields 
\begin{equation}
{\alpha}=gqd
\label{10}
\end{equation}
which yields the following solution for the torsion potential 
${\phi}=e^{-gqdz}$ which displays the coupling between the electric 
charge and the torsion coupling constant.In the time independent case we have
an expression for the electromagnetic vector potential in terms of 
the scalar field as
\begin{equation}
{A}^{0}=\frac{({\nabla}.{S})^{\frac{1}{2}}{( 1+f^{2}{\phi}^{2})}^{\frac{1}{2}}}{g{\phi}}
\label{11}
\end{equation}
From this expression it is very easy to compute the electrostatic field as 
\begin{equation}
E=-{\nabla}A^{0}=-({\nabla}.S)^{\frac{1}{2}}\frac{(1-\frac{1}{2}{\phi})}{g{\phi}^{2}}
\label{12}
\end{equation}
where to simplify matters we have considered that the divergence of the torsion is constant.In the static case the energy can be computed from the Lagrangian above as
\begin{equation}
T_{\mu\nu}=(1+f^{2}){\partial}_{\mu}{\phi}{\partial}_{\nu}{\phi}-{\eta}_{\mu\nu}L
\label{13}
\end{equation}
The energy is
\begin{equation}
{\epsilon}=\frac{1}{4}(E^{2}-B^{2})-\frac{1}{2}g^{2}A^{2}{\phi}^{2}
\label{14}
\end{equation}
In the time dependent case a relation between the angular momentum of the scalar field and torsion is obtained by using the definition of the angular
momentum
\begin{equation}
J_{ij}={\int}(T_{0i}x_{j}-T_{0j}x_{i})d^{4}x
\label{15}
\end{equation}
By making use of equation (\ref{13}) and the definition of the torsion as the gradient of the scalar field yields
\begin{equation}
J_{ij}=(1+f^{2}){\int}S_{0}(S_{i}x_{j}-S_{j}x_{i})d^{4}x
\label{16}
\end{equation}
One must notice that this expression is similar to the relation between spin and torsion which appears in Einstein-Cartan gravity and to the 2+1 Kerr solution discovered by Jackiw \cite{11}. 
It is interesting to note that when the curvature scalar is introduced into the Lagrangian one is able to show that the torsion potential is in fact a massive Higgs field.This can be show by considering the following lagrangian
\begin{equation}
L=R(S)-m^{2}{\phi}^{2}-V({\phi})
\label{17}
\end{equation}
where in considering just the flat metric terms and remiding that the Ricci scalar is given by $R(S)={\partial}S+S^{2}$ where S represents the torsion vector,the Lagrangian (\ref{13}) reduces to
\begin{equation}
L={\partial}^{2}{\phi}-({\partial}{\phi})^{2}-m^{2}{\phi}^{2}-V({\phi})
\label{18}
\end{equation}
The Euler-Lagrange equation for this Lagrangian yields
\begin{equation}
{\partial}_{\mu}{\partial}^{\mu}{\phi}+m^{2}{\phi}=-\frac{{\partial}V}{{\partial}{\phi}}
\label{19}
\end{equation}
Which is the equation for a massive Higgs field given by the torsion potential.Recently in a similar approach Kleinert \cite{10} discussed the spontaneous symmetry breaking for the electroweak bosons induced by the torsion potential,nevertheless in his case the Higgs potential did not coincide with the torsion potential.A more detailed investigation of our model as well a investigation of the Casimir effect in spacetimes with torsion may appear elsewhere.
\section*{Acknowledgements}
I am very much indebt to Professor A.Wang,I.D. Soares and 
H.Kleinert for helpful discussions on the subject of this paper.
Thanks are also due to Universidade do Estado do Rio de Janeiro
(UERJ) for financial Support.

\end{document}